\documentclass[prb,preprint,showpacs,preprintnumbers,amsmath,amssymb]{revtex4-1}

\usepackage[utf8]{inputenc}
\usepackage{amsmath, amssymb,graphicx}
\usepackage[cdot,mediumqspace,squaren]{SIunits}
\usepackage{epsfig}

\renewcommand{\vec}{\mathbf} 
\newcommand{\e}[1]{\text{e}^{#1}} 
\renewcommand{\Re}[1]{\text{Re}\left\{#1\right\}} 
\newcommand{\eps}{ \varepsilon} 
\newcommand{\db}[1]{\overline{\overline{\vec{#1}}}} 

\graphicspath{{./}}

\begin{document}

\title{Wave Instabilities and Unidirectional Light Flow in a Cavity with Rotating Walls}

\author{Sylvain Lanneb\`{e}re\textsuperscript{1}}
\author{M\'{a}rio G. Silveirinha\textsuperscript{1,2}}
\email{To whom correspondence should be addressed:
mario.silveirinha@co.it.pt}
 \affiliation{\textsuperscript{1}
Department of Electrical Engineering, University of Coimbra and
Instituto de Telecomunica\c{c}\~{o}es, 3030-290 Coimbra, Portugal}
\affiliation{\textsuperscript{2}University of Lisbon -- Instituto
Superior T\'ecnico, Department of Electrical Engineering, 1049-001
Lisboa, Portugal}

\date{\today}

\begin{abstract}

We investigate the conditions for the emergence of wave
instabilities in a vacuum cavity delimited by cylindrical metallic
walls under rotation. It is shown that for a small vacuum gap and
for an angular velocity exceeding a certain threshold, the
interactions between the surface plasmon polaritons supported by
each wall give rise to an unstable behavior of the electromagnetic
field manifested in an exponential growth with time. The
instabilities occur only for certain modes of oscillation and are
due to the transformation of kinetic energy into electromagnetic
energy. We also study the possibility of having asymmetric light
flows and optical isolation relying on the relative motion of the
cavity walls.
\end{abstract}

\date{\today}


\maketitle
\section{Introduction}
The interaction of the quantum vacuum electromagnetic fields and
electrically neutral and polarizable macroscopic bodies with rapidly
changing geometry, often referred to as the dynamical Casimir effect
\cite{dodonov_current_2010}, has been extensively studied in the
literature
\cite{fulling_radiation_1976,frolov_excitation_1986,schwinger_casimir_1992,barton_quantum_1993,barton_quantum_1996,eberlein_sonoluminescence_1996,maslovski_casimir_2011,maghrebi_scattering_2013}.
In particular, for bodies in relative translational motion this
interaction is at the origin of the ``quantum friction'' effect
which was predicted to occur when two closely spaced perfectly
smooth parallel surfaces are sheared past one another
\cite{teodorovich_contribution_1978,levitov_van_1989,mkrtchian_interaction_1995,pendry_shearing_1997,volokitin_theory_1999,volokitin_near-field_2007,Milton_Hoye_Brevik,
Horshley, Hoye_Brevik, Hoye_Brevik_2, Hoye_Brevik_3}. The rigorous
physical description of this effect, and in particular its existence
at zero temperature has been the subject of a continued debate
\cite{philbin_no_2009,pendry_quantum_2010,leonhardt_comment_2010,pendry_reply_2010,
Milton_Hoye_Brevik}.

Recently, an important advancement in the physical understanding of
this effect was reported in Refs. \cite{silveirinha_theory_2014,
silveirinha_quantization_2013}, where it was proven that the quantum
friction force emerges even at zero temperature and for lossless
dielectric materials in shear motion with a relative velocity
exceeding twice the Cherenkov threshold. Related ideas have been
developed in parallel using different approaches \cite{ Henkel_2015,
maghrebi_cherenkov}. Furthermore, it was highlighted that in some
scenarios the quantum friction effect has a classical analog and may
be determined by optomechanical interactions that create
electromagnetic wave instabilities \cite{silveirinha_theory_2014,
silveirinha_optical_2014, maslovski_friction_2013}. The
instabilities -- i.e., the natural modes of the system with
amplitude growing with time -- are developed because of the coupling
between the guided modes supported by each surface, and result from
the conversion of kinetic energy into electromagnetic energy, which
is the physical origin of the friction force. The conditions
required for the emergence of the instabilities in planar geometries
were studied in detail in Refs.
\cite{silveirinha_theory_2014,silveirinha_quantization_2013,silveirinha_optical_2014}.
Remarkably, the unstable time evolution of the electromagnetic field
is anchored in a spontaneous parity-time symmetry breaking of the
system and in a phase transition wherein the eigenmodes spectrum
becomes complex valued \cite{silveirinha_spontaneous_2014}.

Interestingly, related instabilities -- known as Kelvin-Helmholtz
instabilities -- may arise when the relative velocity of two fluids
in contact (e.g., the wind blowing over water) exceeds a certain
threshold \cite{chandrasekhar_hydrodynamic_1981}. Moreover,
Kelvin-Helmholtz-type instabilities are well known in plasma physics
and develop in relativistic shear flows of collisionless plasmas in
contact due to the coupling between electron plasma waves mediated
by the electromagnetic field \cite{chandrasekhar_hydrodynamic_1981,alves_large-scale_2012,grismayer_dc-magnetic-field_2013,liang_magnetic_2013,alves_electron-scale_2014,nishikawa_magnetic_2014,alves_transverse_2015}.
Such instabilities are believed to play an important role in
astrophysical scenarios, for example at the interfaces between
astrophysical jets and the interstellar medium. Recently, it was
shown that Kelvin-Helmholtz-type instabilities can develop as well
when there is a vacuum gap in between the sheared plasmas, and this
finding was numerically verified with multidimensional
particle-in-cell simulations \cite{alves_shearing_2015}.

Here, we extend the study of electromagnetic instabilities to
material bodies under rotation. The possibility of wave
amplification by a rotating body was first suggested in the
pioneering work of Zel'Dovich \cite{zeldovich_generation_1971}. More
recently, the spontaneous emission of light by a single rotating
object was studied with the help of the fluctuation dissipation
theorem in the framework of quantum electrodynamics
\cite{manjavacas_vacuum_2010,maghrebi_spontaneous_2012,maghrebi_nonequilibrium_2014}.
Similar to the results of Zel'Dovich, it was shown that light is
emitted only for certain specific modes of oscillation and is
associated with a friction-type torque. The interactions between a
rotating object with a flat metallic surface were investigated in
Ref. \cite{zhao_rotational_2012}, and it was found that the quantum
friction force can be strongly enhanced due to excitation of surface
plasmon polaritons (SPPs).

Contrary to these previous works, the analysis of the present
article relies on simple classical electrodynamics. In particular,
it is highlighted that similar to the case of two bodies in shear
translational motion \cite{silveirinha_optical_2014}, the emergence
of a friction-type force for a rotational motion is deeply rooted in
the development of classical Kelvin-Helmholtz-type instabilities
that lead to the spontaneous conversion of kinetic energy into
electromagnetic energy. To illustrate the ideas, we consider a
simple canonical geometry that corresponds to a vacuum cavity
delimited by two cylindrical metallic walls under rotational
motion. Using non-relativistic classical electrodynamics, we
determine the natural oscillation frequencies of the cavity and find
in which circumstances the moving walls start to spontaneously emit
light. It is shown that an unstable behavior is always accompanied
by the emergence of friction-type mechanical torque that acts to
oppose the relative motion and to stop the instability. Finally, we
study how the relative motion of the walls affects the light
propagation in the cavity, and show that under some conditions it is
possible to have a strongly unidirectional and nonreciprocal light
flow.


\section{Natural modes of the system} \label{sec:natural_modes}
\begin{figure}[!ht]
\centering
\includegraphics[width=.7\linewidth]{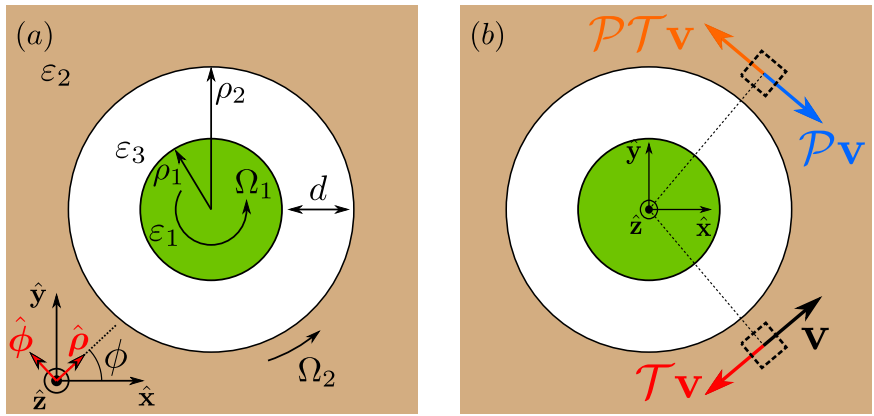}
         \caption{(a) The system under study: a vacuum cavity is
delimited by two cylindrical metallic walls that rotate with angular
velocities $\Omega_1$ and $\Omega_2$. (b) Schematic representing the invariance of the system under $\mathcal{PT}$-symmetry. A generic “element” of the rotating body (delimited by the framed box in the bottom region) is transformed under the time reversal operation as $\mathcal{T} \vec{v}$, under parity operation as $\mathcal{P}\vec{v}$  and under the parity-time operation as $\mathcal{PT}\vec{v}$. Here, the parity operator is taken equal to $\mathcal{P}: \left( x,y,z \right) \to \left(  x, - y, z \right)$ . From an electromagnetic point of view, the parity-time transformation leads to a structure that has the same response as the original one.
}
\label{fig:system_under_study}
\end{figure}

The system under study is a two-dimensional vacuum cavity with
thickness $d$ surrounded by two metallic cylindrical walls rotating
with angular velocities $\Omega_1$, $\Omega_2$ with respect to the
$z$ axis, as depicted in Fig. \ref{fig:system_under_study} (a). The
system is invariant to translations along the $z$-direction. The
metals response is assumed to be determined by the Drude dispersion
model.

The angular velocities of the cylinders are supposed to be time
independent. As discussed in the following, strictly speaking this
condition may require the application of an external force to
counterbalance a friction-type torque due to possible optical
instabilities. In practice, if the metallic walls are sufficiently
massive the effect of the friction torque is expected to be
negligible in the time scale determined by the growth rate of the
electromagnetic fields.

Next, we characterize the cavity's natural modes with complex
oscillation frequencies $\omega= \omega'+i \omega''$. To do so, we
use a purely classical approach to expand the electromagnetic fields
in the different cavity regions in cylindrical harmonics and derive
the characteristic equation by matching the tangential fields at the
interfaces.

It is important to note that different from the case of a body in
uniform translational motion
\cite{silveirinha_theory_2014,silveirinha_quantization_2013,silveirinha_optical_2014},
a body in uniform circular motion is subject to a centripetal
acceleration.
Unfortunately there is no simple description of the electromagnetic
response of a macroscopic medium in accelerated motion. It is well
known that an isotropic uniform (i.e., invariant to translations
along the direction of motion) dielectric medium moving with
constant velocity ${\bf{v}}=v {\hat{\vec{x}}}$ with respect to some
inertial frame (the lab frame) is seen  as a bianisotropic
medium in this reference frame. Indeed, using a relativistic transformation of
the fields it is possible to prove that in the lab frame the
electromagnetic fields are linked as $ \begin{pmatrix} \vec{D} \\
\vec{B} \end{pmatrix}=\begin{pmatrix} \eps_0 \db{\eps} &
\frac{1}{c}\db{\vartheta} \\ \frac{1}{c} \db{\zeta} & \mu_0\db{\mu}
\end{pmatrix} \cdot \begin{pmatrix} \vec{E} \\ \vec{H} \end{pmatrix}
$ with the constitutive parameters given by \cite{Kong_book}
\begin{subequations}
\begin{align}
  \db{\eps} &=\eps_t \left(\db{I} - \hat{\vec{x}}\hat{\vec{x}}\right) + \eps ~\hat{\vec{x}} \hat{\vec{x}}\rm{,} & &\eps_t =\eps\frac{1-\beta^2}{1-n_m^2\beta^2}, \label{E:eqparsA}\\
    \db{\mu} &=\mu_t \left(\db{I} - \hat{\vec{x}}\hat{\vec{x}}\right) + \mu ~\hat{\vec{x}} \hat{\vec{x}}\rm{,} & &\mu_t =\mu\frac{1-\beta^2}{1-n_m^2\beta^2},\\
      \db{\zeta} &=-\db{\vartheta}= -a ~\hat{\vec{x}} \times \db{I}\rm{,} & &a = \beta \frac{n_m^2- 1 }{1-n_m^2\beta^2},
            \label{E:eqparsC}
\end{align}
\end{subequations}
being $\beta=v/c$, and $n_m^2=\eps\mu$, $\eps$ and $\mu$ the
material parameters in the rest (comoving) frame. Equations
\eqref{E:eqparsA}-\eqref{E:eqparsC} are a direct consequence of
formula (76.9) of Ref.\cite{landau_electrodynamics_1984} and their
detailed derivation can be found in Ref.\cite{Kong_book}.  When the
medium is dispersive the parameters $\db{\eps}$, $\db{\mu}$, etc,
must be evaluated at the Doppler shifted frequency
\cite{silveirinha_optical_2014, silveirinha_spontaneous_2014}. It is
relevant to highlight that the material parameters
\eqref{E:eqparsA}-\eqref{E:eqparsC} ensure that the plane wave
natural modes seen in the lab frame have a dispersion that differs
from that seen in the co-moving frame simply by a relativistic
Doppler shift transformation. Unfortunately, formulas
\eqref{E:eqparsA}-\eqref{E:eqparsC} are difficult to generalize to
the case of rotating bodies, mainly because there is no inertial
frame wherein a rotating body is instantaneously at rest.

These features greatly complicate the exact physical
characterization of the wave phenomena in the cylindrical cavity.
Thus, for the sake of simplicity, we will suppose that to a first
approximation the transformed constitutive parameters
\eqref{E:eqparsA}-\eqref{E:eqparsC} are locally valid at each point
of the moving medium. Moreover, we restrict our attention to the
quasi-electrostatic limit and to velocities $v=\Omega\rho$ ($\rho$
is the radial distance to the center of the cavity) small with
respect to the light velocity $c$, so that the bianisotropic nature
of the transformed constitutive parameters can be neglected. Note
that in the quasi-static limit the electric field dominates over the
magnetic field and this justifies the neglect of the crossed
material parameters ($\db{\zeta}, \db{\vartheta}$). Within these
assumptions, equations \eqref{E:eqparsA}-\eqref{E:eqparsC} reduce to
$\db{\eps}\approx \eps $, $\db{\mu}\approx \mu$, $\db{\zeta}
=-\db{\vartheta} \approx 0 $, and the influence of the rotational
motion dwells only in the Doppler shifted frequency
$\tilde{\omega_i}$:
\begin{equation}
\varepsilon _i  = \varepsilon _i \left( {\tilde \omega _i } \right),
\quad \mu _i  = \mu _i \left( {\tilde \omega _i } \right).
\label{E:eqparsDop}
\end{equation}
Here, $\tilde{\omega_i}$ represents the frequency in the frame
instantaneously comoving with the relevant point of the $i$-th
material with velocity ${\bf{v}}_i$. In case of a translational
motion along the $x$-direction, it can be related to the frequency
$\omega$ in the laboratory frame as $\tilde{\omega_i}=\omega - k_x
v_i$ where $k_x  =  - i{\bf{\hat x}} \cdot \nabla$ is the wave
number along the $x$ direction. For a rotational motion
${\bf{v}}_i=\Omega_i \rho {\hat{\vec{\phi}}}$, and thus it follows
that $\tilde \omega _i = \omega + \Omega _i \,i\partial _\phi$,
where $\partial _\phi   =
\partial /\partial \phi $ is the derivative with respect to the
azimuthal angle. The dependence of the constitutive parameters on a
spatial derivative ($\partial _\phi$) is consistent with the fact
that a frequency dispersive medium in motion becomes spatially
dispersive. For waves with an azimuthal variation of the form $\e{ i
n \phi}$ the Doppler shifted frequency is given simply by:
\begin{equation}
 \tilde{\omega_i}=\omega-n\Omega_i,  \label{E:omegatil}
\end{equation}
where $n = 0, \pm 1, ...$ is the azimuthal quantum number. In
summary, in the quasi-static limit and under a non-relativistic
approximation a dielectric under a rotational motion is
characterized in the lab frame by the same material parameters
$\varepsilon_i$ and $\mu_i$ as in the rest frame, but
$\varepsilon_i$ and $\mu_i$ need to be evaluated at the Doppler
shifted frequency $\tilde \omega _i$. In all the examples of the
article the materials do not have a magnetic response ($\mu_i = 1$).
We numerically verified (not shown) that this theory applied to the
case of metal slabs in relative translational motion gives results
consistent with the exact relativistic solution.

Using the proposed formalism it is now a simple task to find the
cavity modes. It is clear that because of the cylindrical symmetry
the modes can be classified according to the azimuthal variation
$\e{ i n \phi}$. Here, we are interested in $p$-polarized modes with
electric field parallel to the $xoy$ plane and magnetic field
directed along the $z$-symmetry axis. An ansatz for the magnetic
field in the lab frame is $H_z =\e{in\phi}f_n(k\rho)$ with $f_n$ a
cylindrical Bessel function of the first kind ($J_{n}$) or of the
second kind ($Y_{n}$). Hence, taking into account the specific
asymptotic conditions to be satisfied in each part of the cavity,
the magnetic field in each region of space of Fig.
\ref{fig:system_under_study} is
\begin{equation}
 H_z(\rho,\phi)= \e{ i n \phi} \left\{
 \begin{tabular}{l l l}
   $ C_{11} J_{|n|}\left(\frac{\omega}{c}\sqrt{\eps_1} \rho\right)$,  & \hspace{0.5cm} & $\rho<\rho_1$
   \\
   $C_{31} J_{|n|}\left(\frac{\omega}{c}\sqrt{\eps_3} \rho\right)+ C_{32} Y_{|n|}\left(\frac{\omega}{c}\sqrt{\eps_3} \rho\right)$, & &  $\rho_2>\rho>\rho_1$
   \\
   $C_{21} H_{|n|}^{(1)}\left(\frac{\omega}{c}\sqrt{\eps_2} \rho\right)  $, & &  $\rho>\rho_2$
 \end{tabular}
\right.
\end{equation}
where $H_{n}^{(1)}=J_{n}+iY_{n}$ is the Hankel function of the first
kind, and $C_{ij}$ are constant coefficients. The permittivity of
each region $\varepsilon_i$ is evaluated at the corresponding
Doppler shifted frequency $\tilde \omega _i$. The azimuthal
component of the electric field in the $i$-th region is given by
\begin{align}
  E_\phi(\rho,\phi) &=  \frac{1}{i \omega \varepsilon_0 \eps_i\left(\tilde \omega _i\right)} \cdot \frac{\partial H_z}{\partial \rho}(\rho,\phi).
\end{align}
By imposing the continuity of $H_z$ and $E_\phi$ at the two
material-vacuum interfaces, one obtains the following 4 $\times$ 4
homogeneous matricial system (here $'$ represents the derivative
with respect to the argument)
\begin{equation}\label{E:matrix_eigenvalues}
 \begin{pmatrix}
 J_{|n|}\left(\frac{\omega}{c}\sqrt{\eps_1} \rho_1\right) & -J_{|n|}\left(\frac{\omega}{c}\sqrt{\eps_3} \rho_1 \right) & -Y_{|n|}\left(\frac{\omega}{c}\sqrt{\eps_3} \rho_1 \right) & 0 \\
   \frac{ 1}{ \sqrt{\eps_1}} J'_{|n|}\left(\frac{\omega}{c}\sqrt{\eps_1} \rho_1\right) & -\frac{ 1}{ \sqrt{\eps_3}} J'_{|n|}\left(\frac{\omega}{c}\sqrt{\eps_3} \rho_1\right) & -\frac{ 1}{ \sqrt{\eps_3}} Y'_{|n|}\left(\frac{\omega}{c}\sqrt{\eps_3} \rho_1\right) & 0 \\
    0 & J_{|n|}\left(\frac{\omega}{c}\sqrt{\eps_3} \rho_2 \right) & Y_{|n|}\left(\frac{\omega}{c}\sqrt{\eps_3} \rho_2\right) &  -H_{|n|}^{(1)}\left(\frac{\omega}{c}\sqrt{\eps_2} \rho_2\right) \\
      0 & \frac{ 1}{ \sqrt{\eps_3}} J'_{|n|}\left(\frac{\omega}{c}\sqrt{\eps_3} \rho_2 \right) & \frac{ 1}{ \sqrt{\eps_3}} Y'_{|n|}\left(\frac{\omega}{c}\sqrt{\eps_3} \rho_2\right) &  -\frac{ 1}{ \sqrt{\eps_2}} H_{|n|}^{\prime (1)}\left(\frac{\omega}{c}\sqrt{\eps_2} \rho_2\right)
 \end{pmatrix}
  \begin{pmatrix}
  C_{11} \\
  C_{31} \\
  C_{32} \\
  C_{21}
 \end{pmatrix}=0
 \end{equation}
whose nontrivial solutions determine the cavity modes. The natural
frequencies of oscillation $\omega = \omega'+i \omega''$ can be
found by setting the determinant of the matrix equal to zero.
Because the Bessel functions lead to a transcendental characteristic
equation, the calculation of the natural frequencies can only be
done with numerical methods. Consistent with the assumptions that
led to Eq. \eqref{E:eqparsDop}, we are interested in subwavelength
metallic cavities for which $k \rho \ll 1$. As shown next, in this
case it is possible to greatly simplify the problem and obtain an
approximate algebraic characteristic equation.

Indeed when $k \rho \ll 1$ the asymptotic form of the cylindrical
Bessel functions can be used
\begin{subequations}
\begin{gather}
J_{|n|}\left(k \rho\right) \underset{k \rho \ll 1}{\longrightarrow} \alpha_1 \left( k \rho \right)^{|n|}, \\
Y_{|n|}\left(k  \rho\right) \underset{k \rho \ll 1}{\longrightarrow}  \alpha_2  \left( k \rho \right)^{-|n|},
\end{gather}
\end{subequations}
where $\alpha_i$ are constant coefficients and it is assumed that $n
\neq 0$. The case $n=0$ is not interesting to us because waves with
a zero azimuthal quantum number do not experience a Doppler shift
($\tilde \omega_i =\omega$ see Eq. \eqref{E:omegatil}), and hence it
is evident that for $n=0$ there are no instabilities. In this
context, the homogeneous matricial system can be rewritten as
\begin{equation}
 \begin{pmatrix}
 - 1&  1&  1 & 0 \\
 -\frac{1}{ \eps_1} & \frac{1}{\eps_3}    & -\frac{1}{\eps_3}  & 0 \\
  0 &  \left(\frac{\rho_2}{\rho_1}\right)^{|n|} &  \left(\frac{\rho_2}{\rho_1}\right)^{-|n|} &  -  1  \\
  0 & \frac{1}{\eps_3 \rho_1}   \left(\frac{\rho_2}{\rho_1}\right)^{|n|-1} & \frac{-1}{\eps_3 \rho_1} \left(\frac{\rho_2}{\rho_1}\right)^{-|n|-1} &  \frac{1}{\eps_2 \rho_2}
 \end{pmatrix}
  \begin{pmatrix}
  A_{11} \\
  A_{31} \\
  A_{32} \\
  A_{21}
 \end{pmatrix}
 =0,
 \end{equation}
where $A_{ij}$ are some constant coefficients. The corresponding
characteristic equation is:
\begin{equation} \label{E:characteristic_equation}
  \frac{ \left[ 1 - \left(\frac{\rho_2}{\rho_1}\right)^{2 |n| }\right] }{ \left[ 1 + \left(\frac{\rho_2}{\rho_1}\right)^{2 |n| }\right] }    = \frac{ \eps_1(\tilde{\omega}_1) \eps_3 (\tilde{\omega}_3) + \eps_2(\tilde{\omega}_2) \eps_3 (\tilde{\omega}_3) }{ \eps_1(\tilde{\omega}_1) \eps_2(\tilde{\omega}_2) + \eps_3^2(\tilde{\omega}_3)
  }.
\end{equation}
Clearly, in this quasi-static approximation [Eq.
\eqref{E:characteristic_equation}] the natural oscillation
frequencies $\omega = \omega'+i \omega''$ only depend on the ratio
between the radii but not on the specific values of the individual
radii. The impact of a finite cavity radius can be investigated by
directly solving equation \eqref{E:matrix_eigenvalues}. It is
important to highlight that Eq. \eqref{E:characteristic_equation} is
 nothing more than -- apart from the Doppler-shifted frequencies in the material
parameters -- the dispersion of the cavity modes in the quasi-static
limit. This observation makes clear that our analysis can be
extended in a straightforward manner to other geometries. In the
rest of the article, it is assumed that the middle layer is a vacuum
($\eps_3=1$), and that the other two materials are modeled by a
Drude dispersion model $\eps_i(\tilde{\omega}_i)= 1-
\omega_p^2/\left[\tilde{\omega}_i( \tilde{\omega}_i + i \Gamma )
\right]$, being $\omega_p$ the plasma frequency and $\Gamma$ the
collision frequency.

\subsection{Instabilities in the quasi-static limit}

\begin{figure}[!ht]
\centering
\includegraphics[width=1\linewidth]{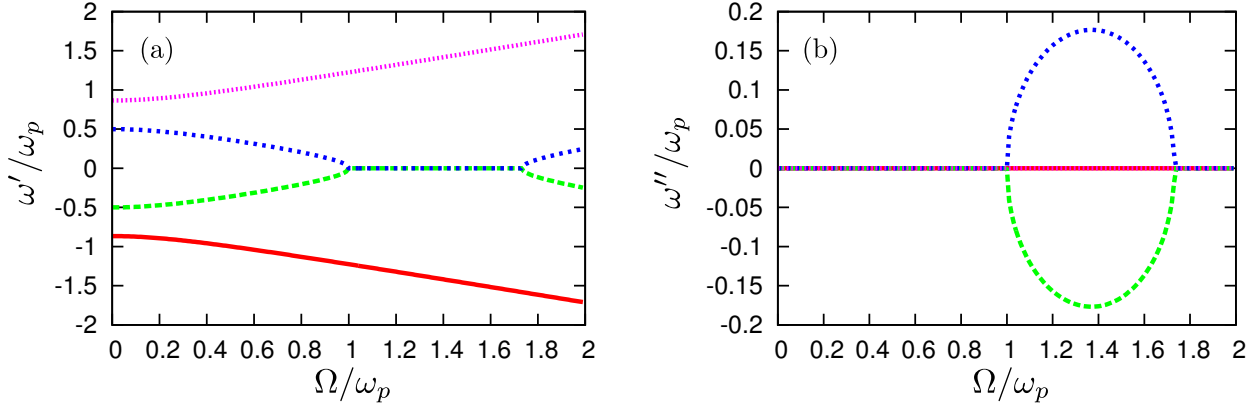}
         \caption{Free oscillation complex frequencies of the
  cavity $\omega= \omega'+i \omega''$ for $\Omega_1=-\Omega_2=\Omega/2$, $n=1$, $\rho_2/\rho_1=2$ and $\Gamma=0$. (a) $\omega'/\omega_p$ as a function of $\Omega/\omega_p$. (b) $\omega''/\omega_p$ as a   function of $\Omega/\omega_p$.}
\label{fig:natural_modes_vs_velocity}
\end{figure}

In the following, we use the algebraic characteristic equation
\eqref{E:characteristic_equation} to characterize the cavity natural
frequencies. Figure \ref{fig:natural_modes_vs_velocity} represents
the calculated $\omega$ as a function of the relative angular
velocity $\Omega$, for $\Omega_1=-\Omega_2=\Omega/2$, $n=1$,
$\rho_2/\rho_1=2$ and no material loss ($\Gamma=0$). As seen, for
angular velocities greater than a threshold velocity approximately
coincident with the plasma frequency of the metal $\omega_p$, two of
the eigenwaves have complex oscillation frequencies. In particular,
one of the modes has $\omega''$ with a positive imaginary part
corresponding to waves growing with time as $e^{\omega ''t}$, i.e.
to an unstable system.

Interestingly, the plot of $\omega''$ versus the angular frequency
is symmetric with respect to the horizontal axis, such that the
complex frequencies occur in pairs $\omega' \pm i \omega''$. This
feature is characteristic of systems with a broken parity-time
(${\cal P}{\cal T}$) symmetry
\cite{bender_pt-symmetric_1999,bender_making_2007,silveirinha_spontaneous_2014}.
Indeed, in case of lossless materials the system of Fig.
\ref{fig:system_under_study} (a) is invariant under the ${\cal
P}{\cal T}$ operation, being the parity operation understood as the
transformation $\left( {x,y,z} \right) \to \left( { x, - y, z}
\right)$ (one could as well choose the transformation $\left(
{x,y,z} \right) \to \left( { -x, y, z} \right)$). Note that as
illustrated in Fig. \ref{fig:system_under_study} (b), the
time-reversal operator flips the velocity of the medium
\cite{silveirinha_spontaneous_2014}, while the parity operator flips
the $y$-component of the velocity, and hence the combined ${\cal
P}{\cal T}$ operation only flips the $x$-component of the velocity,
as it should be so that the medium stays invariant under a
coordinate transformation of the form $\left( {x,y,z} \right) \to
\left( { x, - y, z} \right)$. Thus, for a lossless system the
emergence of system instabilities is a manifestation of a broken
${\cal P}{\cal T}$-symmetry similar to the planar case studied in
Ref. \cite{silveirinha_spontaneous_2014}, and implies that the time
evolution of electromagnetic waves is described by a non-Hermitian
${\cal P}{\cal T}$-symmetric operator. Unstable natural modes are
not invariant under the ${\cal P}{\cal T}$ operation, even though
the physical system has that symmetry.

In Fig. \ref{fig:system_under_study} the instabilities have a vanishing real part $\omega'$ and correspond to a static field, predominantly electric, growing exponentially with time. This feature is
specific to the scenario where both cylinders rotate in opposite
directions with the same angular velocity.

When the angular velocities of the two materials are asymmetric
($\Omega _1  + \Omega _2 \ne 0$) the ratio between the amplitudes of
the magnetic and electric fields grows with $\Omega _1  + \Omega
_2$, and the frequency $\omega'$ is transformed as:
\begin{equation}
\omega ' \to \omega '\left(\Omega \right) + n\frac{{\Omega _1  +
\Omega _2 }}{2}
\end{equation}
where  $\Omega \equiv \Omega _1  - \Omega _2$ and $\omega''$ remains
invariant. The peak value of $\omega''$ occurs roughly for $\Omega =
1.4 \omega_{p} \approx 2 \omega_{sp}$, being
$\omega_{sp}=\omega_{p}/\sqrt 2$ the surface plasmon resonance. In
particular, when one of the material regions is at rest (let's say
$\Omega_2=0$) one gets $\omega' \gg \omega''$, and the peak
instability is associated with $\omega' \approx \omega_{sp}$ for the
$n=1$ mode. Independent of the value of $\Omega _1 + \Omega _2$, the
amplification occurs only in a finite range of frequencies $\omega'$
in agreement with the conclusions of refs.
\cite{zeldovich_generation_1971,manjavacas_vacuum_2010,maghrebi_spontaneous_2012,maghrebi_nonequilibrium_2014}.

\begin{figure*}[!ht]
\centering
\includegraphics[width=.85\linewidth]{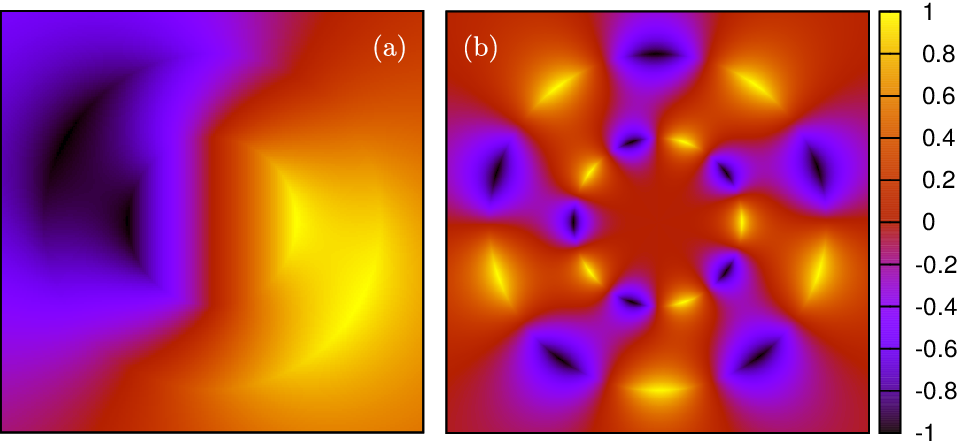}
         \caption{Time snapshot of the real part of the magnetic field $H_z$ associated with the instability for $\Omega_1=-\Omega_2=\Omega/2$ and $\rho_2/\rho_1=2$. (a) $n=1$ and $\Omega/\omega_p=1.37$. (b) $n=5$ and $\Omega/\omega_p=0.28$.}
\label{fig:snapshot_field}
\end{figure*}
A density plot of the magnetic field  associated with the strongest
instability is depicted in Fig. \ref{fig:snapshot_field}(a). The
field has a dipolar symmetry ($n=1$) and  the regions of strongest
field intensity are concentrated close to the surface of each
cylinder, facing each other. This observation together with the fact
that the instabilities are developed for $\Omega \approx 2
\omega_{sp}$ demonstrates that the field exponential growth with
time is due to the interaction of surface plasmon polaritons in each
cylinder, similar to what happens in the planar case
\cite{silveirinha_quantization_2013, silveirinha_optical_2014,
silveirinha_theory_2014, silveirinha_spontaneous_2014}. The field
exponential growth can be pictured as being due to the interaction
of positive frequency harmonic oscillators and negative frequency
harmonic oscillators, such that $\tilde \omega_1$ and $\tilde
\omega_2$ have opposite signs \cite{silveirinha_quantization_2013,
silveirinha_optical_2014}. This interaction is made possible by the
relative rotation of the two cylinders. Oscillators associated with
negative frequencies behave as energy reservoirs that may serve to
pump the oscillations of the system and generate the unstable
behavior \cite{frolov_excitation_1986,silveirinha_quantization_2013,
silveirinha_optical_2014, Horshley2, Jacob}. For example, when
$\Omega_2=0$ it can be checked that $\tilde \omega_1<0$ and $\tilde
\omega_2>0$  in the entire instability region, and thus the moving
region may be regarded as the energy reservoir. In addition, the
plasmonic nature of the interaction can be seen by noting that the
peak instability occurs for $\tilde \omega_1\approx -\omega_{sp}$
and $\tilde \omega_2 \approx \omega_{sp}$. It is instructive to note
that for a translational motion an unstable behavior requires the
coupling of two oscillators (interaction of two different bodies in
relative motion) because with a single moving body it is always
possible to switch to a frame where the body is at rest, and where
evidently instabilities are impossible to occur
\cite{silveirinha_quantization_2013}. In principle, for a rotational
motion the instabilities may occur even with a single body because a
rotating body is in motion in \emph{any} inertial frame. However, if
the velocity of rotation is sufficiently small, as implicit in our
theory, the interaction between different parts of the same body is
ineffective, and similar to the case of a translational motion two
interacting bodies (two different oscillators) are required to
trigger the instability. Indeed, for a single rotating cylinder
surrounded by a vacuum (the limit $\rho_2 \to \infty$ and
$\eps_3=1$) Eq. \eqref{E:characteristic_equation} reduces to
$\eps_1(\tilde{\omega}_1)+1=0$, which evidently does not lead to any
instability.

\begin{figure}[!ht]
\centering
\includegraphics[width=.45\linewidth]{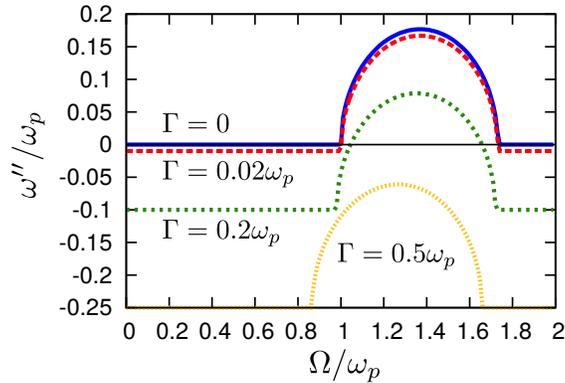}
         \caption{Influence of the metal damping (collision) frequency $\Gamma$ on $\omega''$ for the mode
associated with the unstable behavior for
$\Omega_1=-\Omega_2=\Omega/2$, $n=1$ and $\rho_2/\rho_1=2$.}
\label{fig:natural_modes_vs_loss}
\end{figure}
It is interesting to see how the material loss affects the natural
oscillation frequencies, and in particular whether the instabilities
withstand realistic plasmonic loss. This study is reported in Fig.
\ref{fig:natural_modes_vs_loss} which depicts the imaginary part of
the free oscillation frequency (for the mode with positive
$\omega''$) as a function of the normalized angular velocity. The
effect of damping is roughly equivalent to adding a negative
constant imaginary part to the value of $\omega''$ that reduces the
strength of the growth rate. It is relevant to mention that in the
presence of material loss the system is not anymore ${\cal P}{\cal
T}$-symmetric, and hence the frequency spectrum does not have the
complex conjugation symmetry as in Fig.
\ref{fig:natural_modes_vs_velocity}. The range of angular velocities
$\Omega$ for which $\omega''>0$ becomes narrower with increasing
$\Gamma$, up to a point wherein all the oscillations are damped
($\omega''<0$) and the instability ceases. Importantly, the unstable
behavior is quite robust to the effect of material loss and is
observed even for collision frequencies much larger than those
characteristic of realistic metals ($0.01 < \Gamma /\omega _p  <
0.2$). Furthermore, since the instabilities result from the
hybridization of evanescent waves attached to the individual
cylinders, they are strongly dependent on the value of
$\rho_2/\rho_1$ and the material loss may be partially compensated
by narrowing the vacuum gap.
\begin{figure}[!ht]
\centering
\includegraphics[width=.45\linewidth]{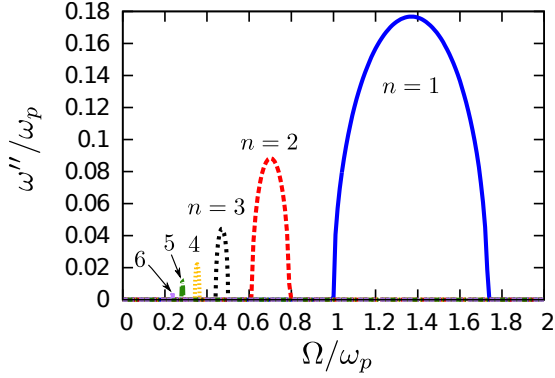}
         \caption{Evolution of $\omega''$ with the azimuthal quantum number $n$ for the mode
associated with the unstable behavior and
$\Omega_1=-\Omega_2=\Omega/2$, $\rho_2/\rho_1=2$ and $\Gamma=0$.}
\label{fig:natural_modes_vs_n}
\end{figure}

The influence of the quantum azimuthal number $n$ on the strength of
the instabilities is investigated in Fig.
\ref{fig:natural_modes_vs_n} for the case
$\Omega_1=-\Omega_2=\Omega/2$. It is seen that the strength of the
instabilities progressively decreases with $n$ and is negligible for
high $n$. This can be understood noting that modes with a large $n$
result from the hybridization of tightly confined surface plasmons
that overlap weakly in the vacuum gap as illustrated in Fig. \ref{fig:snapshot_field}(b) for $n=5$. Stronger instabilities are
obtained for a smaller gap. Remarkably, the threshold angular
velocity for the unstable behavior scales roughly as $1/n$ and hence
decreases with the azimuthal quantum number.
In particular, the angular velocity for which the unstable behavior
is stronger is $\Omega \approx 2 \omega_{sp}/n$. In case one of the
bodies is at rest (e.g., $\Omega_2=0$) the peak instability is
always associated with the frequency $\omega' \approx \omega_{sp}$,
independent of the value of $n$.

Due to the reality of the electromagnetic field, the spectrum
associated with negative values of $n$ is linked to the spectrum
associated with positive values of $n$ as $\omega \to -\omega^*$,
such that the real part of the frequency is flipped, while the
imaginary part is unchanged. In the case wherein $\Omega_2=0$, the
unstable modes with $\omega'>0$ occur for azimuthal quantum numbers
$n$ with the same sign as $\Omega_1$. An intuitive explanation is
that the spontaneous light emission by a rotating body favors
physical channels associated with angular variations in the
direction determined by the moving body.

\subsection{Effect of time retardation}

\begin{figure*}[!ht]
\centering
\includegraphics[width=.95\linewidth]{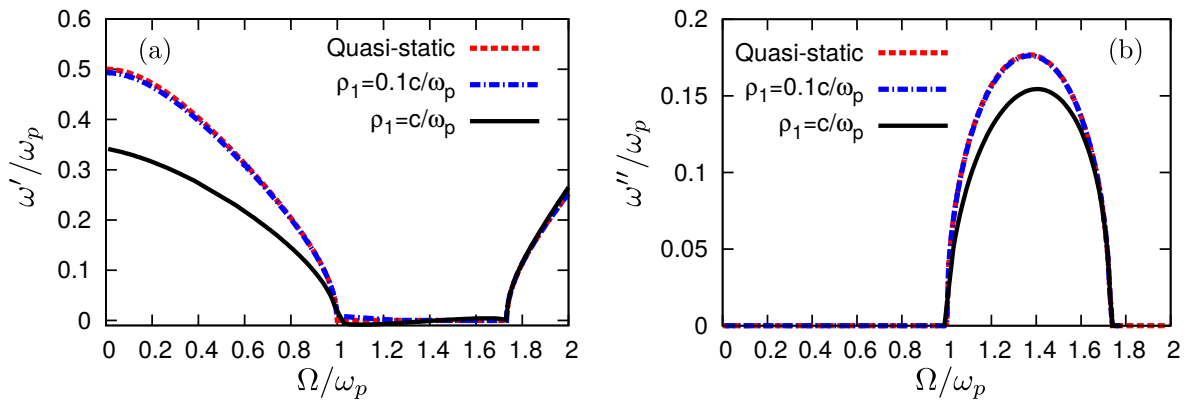}
         \caption{Comparison between the ``exact'' ($\rho_1=c/\omega_p$ and $\rho_1=0.1 c/\omega_p$) and the quasi-static free oscillation complex frequencies of the
  cavity $\omega= \omega'+i \omega''$ for the mode with positive imaginary part. Here $n=1$, $\rho_2/\rho_1=2$ and $\Gamma=0$. (a) $\omega'/\omega_p$ as a function of $\Omega/\omega_p$. (b) $\omega''/\omega_p$ as a   function of $\Omega/\omega_p$.}
\label{fig:compa_quasistatic_normal}
\end{figure*}

It is important to have some idea of the impact of time retardation
effects. Figure \ref{fig:compa_quasistatic_normal} shows a
comparison between the quasi-static theory and the results obtained
by setting the determinant of the matrix in Eq.
\eqref{E:matrix_eigenvalues} equal to zero. In this plot, the
structural parameters are as in the previous section, namely
$\rho_2/\rho_1=2$, and only the mode $n=1$ with positive imaginary
part is represented. As seen, as soon as the radii of the cylinders
become of the order of $c/\omega_p$, the quasi-static approximation
breaks down and the time retardation effects play some role. The
effect of time retardation is to reduce the strength of the
instabilities. The threshold for the emergence of instabilities is
unaffected by the time retardation.

\subsection{Wave energy}

\begin{figure}[!ht]
\centering
\includegraphics[width=.55\linewidth]{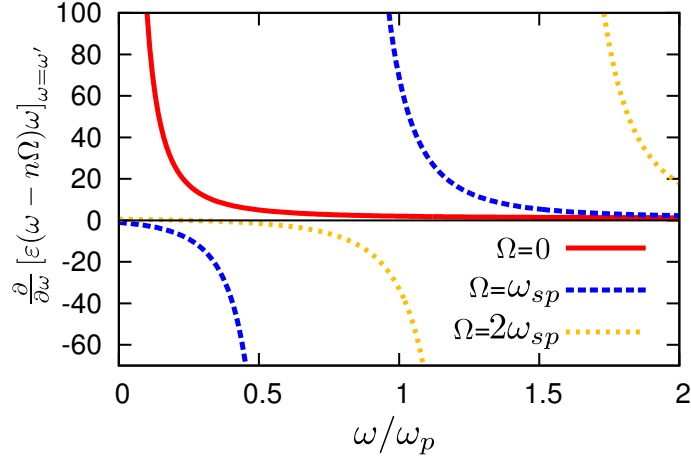}
         \caption{Plot of $ \frac{\partial}{\partial \omega} \left[\eps(\omega-n \Omega) \omega \right]_{\omega=\omega'}$
         as a function of the normalized frequency for $\Omega=0$, $ \omega_{sp}$ and $2 \omega_{sp}$ and a dipolar interaction ($n$=1).
         The function is negative for $0<\omega<\Omega$ .}
\label{fig:wave-energy}
\end{figure}

From the Poynting theorem in the time domain, it is simple to show that for a closed
cavity and in the absence of external current sources the quantity $
E_w=\int^t dt \int\limits_\text{all space} dV \left( \vec{E} \cdot
\frac{\partial \vec{D}}{\partial t} + \vec{H} \cdot \frac{\partial
\vec{B}}{\partial t}\right)$   must be time independent. It is
implicitly assumed that the angular velocity of the rotating bodies
is time-invariant. For lossless materials $E_w$  can be identified
with the total wave energy of the system, which includes the
electromagnetic energy and part of the energy associated with the
kinetic degrees of freedom of the system (see
Ref.\cite{silveirinha_theory_2014} for a detailed discussion). For
time-harmonic fields $\vec{E} = \Re{ \vec{E}_\omega \e{-i\omega
t}}$, $\vec{D} = \Re{\vec{D}_\omega \e{-i\omega t}}$, etc, it is
straightforward to check that $E_w= E_{w0}(t) \e{2\omega'' t}$ where
$E_{w0}(t)$ is a periodic function of time with period
$\pi/\omega'$. The time averaged value of $E_{w0}(t)$ is given by
$E_
{w0,\text{av}}= \int\limits_\text{all space} W_\text{av} ~dV$  where
\begin{align}
W_\text{av} = \frac{1}{4} \Re{\vec{D}_\omega \cdot
\vec{E}_\omega^\ast  \frac{\omega}{i \omega''}} + \frac{1}{4}
\Re{\vec{B}_\omega \cdot \vec{H}_\omega^\ast  \frac{\omega}{i
\omega''}}
\end{align}
is the time-averaged wave energy density envelope. For a weak
instability $\omega'' \ll \omega'$   (limit $\omega'' \to 0$) and
within the approximations implicit in Eq. \eqref{E:eqparsDop} the
wave energy density reduces to $W_\text{av} = \frac{1}{4}
\vec{E}_\omega \cdot \vec{E}_\omega^\ast \frac{\partial}{\partial
\omega} \left[\eps \omega \right] + \frac{1}{4} \vec{H}_\omega \cdot
\vec{H}_\omega^\ast \frac{\partial}{\partial \omega} \left[\mu
\omega \right] $, consistent with a well-known electromagnetic
theory result \cite{landau_electrodynamics_1984}. In the
quasi-static limit the field is predominantly electric and hence:
\begin{align}
 W_\text{av} &\approx \frac{1}{4} \vec{E}_\omega \cdot \vec{E}_\omega^\ast \frac{\partial}{\partial \omega} \left[\eps \omega \right]_{\omega=\omega'} \nonumber \\
 &= \frac{1}{4} \vec{E}_\omega \cdot \vec{E}_\omega^\ast \frac{\partial}{\partial \omega} \left[\eps(\vec{r},\omega-n \Omega) \omega
 \right]_{\omega=\omega'}.
\end{align}
In Figure \ref{fig:wave-energy} we plot $ \frac{\partial}{\partial
\omega} \left[\eps(\omega-n \Omega) \omega \right]_{\omega=\omega'}$
for a Drude plasma as a function of frequency for different values of
the angular velocity and assuming a dipolar mode ($n=1$).
As seen, for nonzero angular velocities $
\frac{\partial}{\partial \omega} \left[\eps(\omega-n \Omega) \omega
\right]_{\omega=\omega'}$ can become negative, i.e., the wave energy
density can be negative in a moving region. In that case, an
increase of the electric field intensity leads to a decrease of the
local wave energy. These findings are completely consistent with the
results of
Refs.\cite{silveirinha_quantization_2013,silveirinha_theory_2014,
Horshley}, where it was shown that the wave energy density has no
lower bound in a material that moves with a velocity that exceeds
the Cherenkov limit. This theory provides an intuitive explanation
of the reason why the instabilities are associated with two
oscillators with oppositely signed frequencies. Indeed, in case of
an unstable behavior (growing exponential) the total wave energy of
the system can be time-independent only if $E_{w0,\text{av}}=0$ .
Thus, an oscillator with positive frequency is required to increase
its wave energy in the exact same proportion as the oscillator with
negative frequency decreases the energy, so that the total wave
energy remains time independent. Note that even though the total wave energy is time independent,
the wave energy density varies with time and in space due to the energy transfer between the two bodies under rotation.

\section{Torque}
An obvious question is: what is the mechanism that pumps the
unstable modes? Is it rooted, similarly to the planar case
\cite{silveirinha_theory_2014}, in the conversion of kinetic energy
into electromagnetic energy? Next, it is confirmed that this is
indeed the case, and to demonstrate such a property we determine the
mechanical torque induced by the instabilities.

The instantaneous force per unit of volume induced by the electromagnetic field
is \cite{stratton_electromagnetic_2007}
\begin{equation}
\vec{f}= \nabla \cdot \db{T} - \frac{\partial}{\partial t}
\vec{g_{\rm{EM}}},
\end{equation}
where $\vec{g_{\rm{EM}}}$ is the electromagnetic momentum density
and $\db{T}$ the Maxwell stress tensor. The torque
$\boldsymbol{\tau}$ resulting from the force $\vec{f}$ is then given
by
\begin{equation}
\boldsymbol{\tau} = \int \vec{r} \times \vec{f}~d^3\vec{r} = \int
\vec{r} \times \left( \nabla \cdot \db{T} - \frac{\partial}{\partial
t} \vec{g_{\rm{EM}}} \right) ~d^3\vec{r} .
\end{equation}
where $\vec{r}$ is the position vector. Integrating by parts, it is
possible to write the contribution of the stress-tensor as an
integral over the surface of the relevant body (at the air side):
\begin{equation}
\boldsymbol{\tau} = \int {{\bf{r}} \times \left( {{\boldsymbol{\hat
\nu }} \cdot \overline{\overline {\bf{T}}} } \right)\,} ds - \int
\vec{r} \times \frac{\partial}{\partial t} \vec{g_{\rm{EM}}}
 ~d^3\vec{r} .
\end{equation}
Here, $\boldsymbol{\hat \nu }$ is a unit vector oriented towards the
outside of the body, and we use the fact that $\sum\limits_i
{{\bf{\hat u}}_i  \cdot \overline{\overline {\bf{T}}} \times
{\bf{\hat u}}_i } = 0$, being ${\bf{\hat u}}_i$ a generic unit
vector along the Cartesian coordinate axes, because the stress
tensor is symmetric. Because in the surface integral the
stress-tensor is evaluated at the air side of the interface we can
use the standard formula
\begin{equation}
\db{T}=\eps_0 \vec{E}\otimes\vec{E} + \mu_0 \vec{H}\otimes\vec{H} -
1/2\left(\eps_0 |\vec{E}|^2 + \mu_0 |\vec{H}|^2\right) \db{I}.
\end{equation}
Clearly, in time harmonic regime the torque (which is a quadratic
function of the electromagnetic fields) grows exponentially as $
e^{2\omega ''t}$. Hence, we can write $\boldsymbol{\tau} =
\boldsymbol{\tau}_0 \left( t \right)e^{2\omega ''t}$, being
$\boldsymbol{\tau}_0$ the envelope of the torque which typically has
a component that oscillates in time with frequency $2\omega '$. The
definition of the electromagnetic momentum density in material media
is surrounded by a century-old controversy
\cite{pfeifer_textbf_2007,barnett_resolution_2010}. We can avoid
this controversy by calculating time-averaged torque, obtained by
time-averaging the envelope of the torque: $\left\langle
{\boldsymbol{\tau}} \right\rangle  =
{\boldsymbol{\tau}}_{0,{\rm{av}}} e^{2\omega ''t}$. It is simple to
check that with this definition one has $\left\langle \frac{\partial
\vec{g_{\rm{EM}}}}{\partial t} \right\rangle =  0$, and hence it is
finally found that:
\begin{equation}
{\boldsymbol{\tau}}_{0,{\rm{av}}}  =
\frac{1}{2}{\mathop{\rm Re}\nolimits} \left\{ {\int {{\bf{r}} \times
\left( {{\boldsymbol{\hat \nu }} \cdot \overline{\overline {\bf{T}}}
_c } \right)\,} ds} \right\}
\end{equation}
where $\db{T}_c=\eps_0 \vec{E}_\omega \otimes\vec{E}_\omega^* + \mu_0
\vec{H}_\omega \otimes\vec{H}_\omega^* - 1/2\left(\eps_0 |\vec{E}_\omega |^2 + \mu_0
|\vec{H}_\omega|^2\right) \db{I}$ is a complex stress tensor written in
terms of the complex vector field amplitudes.

Let us apply this theory to the scenario considered in section
\ref{sec:natural_modes} wherein the material 2 is at rest
($\Omega_2=0$) and the material 1 rotates with angular velocity
$\Omega_1=\Omega$. Straightforward calculations show that the
time-averaged torque per unit of length is
\begin{align}
\frac{\left\langle \boldsymbol{\tau} \right\rangle}{h}  &=
 2\pi \rho_1^2 \frac{\eps_0}{2} {\mathop{\rm
Re}\nolimits} \left\{ {E_{\omega,\rho}(\rho_1)
E_{\omega,\theta}^*(\rho_1)} \right\} e^ {2\omega'' t} ~
\hat{\vec{z}},
\end{align}
where $h$ is the height of the cylinder. This formula shows that the
torque can be roughly estimated as $\left\langle \boldsymbol{\tau}
\right\rangle \sim {\boldsymbol{\varepsilon }}_E \e{2\omega'' t}$
where ${\boldsymbol{\varepsilon }}_E$ is the electric energy stored
in the air cavity at time $t=0$. Hence, in the time scale determined by
$1/\omega''$ the mechanical torque is typically small, and becomes
relevant only when $t \gg 1/\omega''$ due to the exponential growth.

\begin{figure}[!ht]
\centering
\includegraphics[width=.45\linewidth]{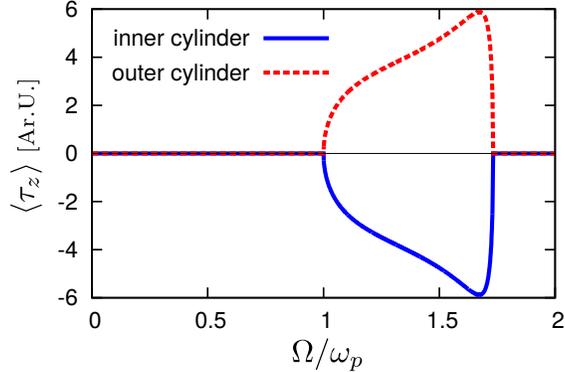}
         \caption{Normalized time averaged torque (in arbitrary units) acting on the inner and outer cylinders as a function of the
          angular velocity for $\rho_2/\rho_1=2$, $\Omega_1=\Omega$, $\Omega_2=0$ and lossless plasmonic materials $\Gamma=0$.}
\label{fig:torque}
\end{figure}

The time-averaged torque calculated with the quasi-static
approximation for the mode with positive imaginary part (the
unstable mode) is represented in Fig. \ref{fig:torque} as a function
of the angular velocity at a given instant of time.
In the numerical simulation it is assumed that only the inner
cylinder is rotating and that $n=1$, $\rho_2/\rho_1=2$ and
$\Gamma=0$. As seen in section \ref{sec:natural_modes}, in this
situation the instabilities are developed for angular velocities
larger than $\omega_p$ due to the interaction of the SPPs supported
by each surface and associated with $\tilde \omega_1 < 0$ and
$\tilde \omega_2 >0$. By comparing Figs.
\ref{fig:natural_modes_vs_velocity} and \ref{fig:torque}, it is seen
that the torque is intimately linked to these instabilities since it
is non zero only for the angular velocities for which the system is
unstable. Importantly, the torque acting on the inner-cylinder is
negative and therefore acts against the rotational motion: it is a
friction-type torque. The torque acting on the outer-cylinder is
exactly the opposite and tends to drag the outer-cylinder into
motion, thereby reducing the relative angular velocity between the
cylinders. This result confirms that the emergence of instabilities
in the system has its origin in the conversion of kinetic energy
into electromagnetic energy. Hence, as mentioned in the
introduction, to keep the angular velocity constant a positive
torque needs to be applied to the inner cylinder in order to
counterbalance this friction-type torque. The results of this
section further highlight the intimate connection between
electromagnetic friction and the coupling between oscillators with
positive and negative frequencies in agreement with previous studies
\cite{Horshley,silveirinha_quantization_2013,silveirinha_optical_2014,silveirinha_theory_2014}.

\section{Unidirectional light flow}

A moving medium is not invariant under a time-reversal
transformation, and consequently it is characterized by a
nonreciprocal electromagnetic response. This property raises
interesting possibilities in the context of asymmetric light flows,
which are otherwise generally forbidden in conventional reciprocal
media (e.g., isotropic dielectrics and metals at rest)
\cite{jalas_what_2013}.

\begin{figure}[!ht]
\centering
\includegraphics[width=.9\linewidth]{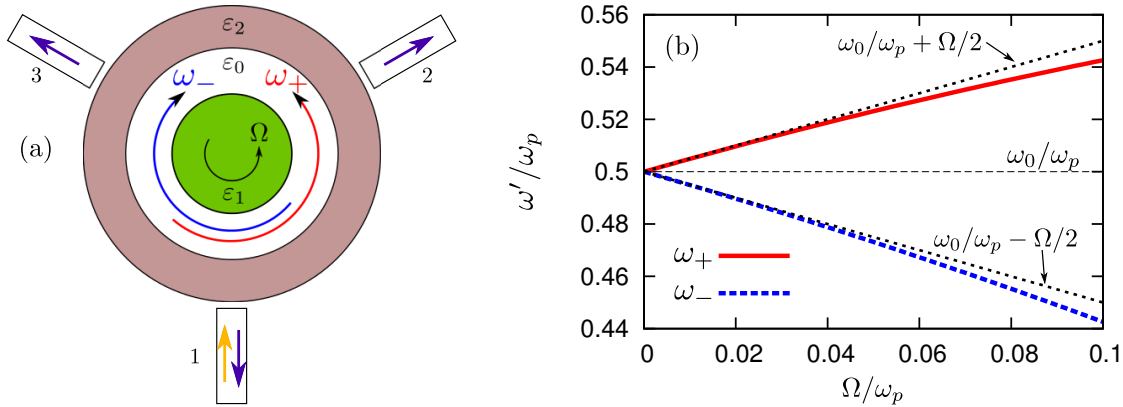}
         \caption{(a) Schematic of the optical circulator with the incoming signal represented by the orange (light gray) arrow and the outgoing waves by the purple (dark gray) arrows. (b) Modes of the cylindrical cavity as a function of the normalized angular velocity for $\rho_2/\rho_1=2$, $\Omega_1=\Omega$, $\Omega_2=0$ and $\Gamma=0.01 \cdot \omega_p$.}
\label{fig:modes_optical_circulator}
\end{figure}

An intuitive picture of the effect of motion is that the phase
velocity depends if the wave propagates downstream or upstream with
respect to the flow of matter
\cite{Stas_Cherenkov,philbin_fiber-optical_2008}. As discussed next,
this feature can be explored to design a light ``circulator''. A
circulator is a nonreciprocal three-port network, that is ubiquitous
in microwave technology \cite{pozar_microwave_2011}. Circulators are
also useful at optical frequencies for optical switching and for the
optical isolation of the light source from the optical channel.
\cite{potton_reciprocity_2004,jalas_what_2013}.  A circulator
ideally only allows transmission from port 1 to 3 or from port 3 to
2 or from port 2 to 1 (see Fig. \ref{fig:modes_optical_circulator};
in the figure it is implicit that optical waveguides are connected
to the cylindrical cavity at the relevant ports). The propagation in
the opposite azimuthal direction (e.g., from port 1 to port 2) is
forbidden.

In conventional circulators, the unidirectional light flow relies on
the existence of constructive or destructive interference patterns
at each output, arising from the different oscillation frequency of
the modes circulating in the clockwise or anti-clockwise directions
in the cavity. Usually the mode splitting is achieved due to a
Zeeman-type effect when magnetic materials (e.g., ferrites and some
garnets \cite{pozar_microwave_2011}) are biased by a static magnetic
field. Because of the drawbacks in terms of size and weight of
permanent magnets, different approaches free of magnetic elements
have been recently proposed
\cite{kodera_artificial_2011,sounas_electromagnetic_2013,sounas_giant_2013,sounas_angular-momentum-biased_2014,estep_magnetic-free_2014,estep_magnetless_2016}.

In what follows, we theoretically demonstrate a new paradigm for an
optical circulator based on a cavity with rotating walls. A related
result was recently reported for the acoustic case
\cite{fleury_sound_2014}. Remarkably, this effect is neither based
on wave instabilities in the cavity nor on the excitation of SPPs in
each slab, but rather on the modal asymmetry between the waves
propagating in the direction of rotation ($n=+1$) and in the counter
propagating direction ($n=-1$)
\cite{wang_magneto-optical_2005,fleury_sound_2014,sounas_giant_2013,sounas_angular-momentum-biased_2014,estep_magnetic-free_2014,fleury_subwavelength_2015,estep_magnetless_2016}.
Indeed, it can be shown using equation
\eqref{E:characteristic_equation} that at small velocities ($\Omega
\ll \omega_{p}$) one of the resonant frequencies of the cavity is
approximately given by
\begin{equation} \label{E:approx_omega_n}
  \omega_n \approx \omega_0 + \frac{n \Omega}{2} - i \frac{ \Gamma}{2}  ,
 \end{equation}
 where $\omega_0 = \omega_{sp} \sqrt{1 - \left(\frac{\rho_1}{\rho_2} \right)^{\left| n\right| } - \frac{\Gamma^2}{2 \omega_p^2}}$ is the resonant frequency of the cavity when the walls are at rest. Then it follows that the natural frequency of oscillation for the modes associated with $n=\pm1$ linearly split as
\begin{equation}
  \omega_\pm \approx \omega_0 \pm \frac{\Omega}{2}.
 \end{equation}
This behavior is illustrated in Fig.
\ref{fig:modes_optical_circulator} where the evolution of the
natural frequencies with the normalized angular velocity is
represented  for a resonator with $\rho_2/\rho_1=2$,
$\Omega_1=\Omega$, $\Omega_2=0$ and $\Gamma=0.01 \cdot \omega_p$. As
seen, the motion of the inner cylinder induces a splitting of the
modes, linear in $\Omega$ at small angular velocities, and somewhat
analogous to the Zeeman splitting obtained with magnetic materials.
The splitting can be understood noting that the resonance condition
is roughly of the form $\omega l/v_p = 2 \pi$, where $l \approx \pi
(\rho_1+\rho_2)$ is the mean perimeter of the cavity and $v_p$ is
the SPP phase velocity in the cavity, which depends on the azimuthal
quantum number. Because of the Fresnel drag one may expect that when
$\Omega>0$ the velocity $v_p$ is larger for modes associated with
positive $n$ as compared to a mode with index $-n$. Thus, this
indicates that $\omega_+>\omega_-$, consistent with the analytical
model.

When the circulator is excited in port 1, the transmissivities for
ports 2 and 3 depend on the frequencies of the cavity modes as
follows (the formula below corrects a typo in Ref.
\cite{fleury_sound_2014})
\cite{wang_magneto-optical_2005,fleury_sound_2014}
\begin{subequations}
\begin{align}
T_{1\to2} &=   \left|  \frac{2}{3} \left( \frac{ \e{-i\frac{2\pi}{3}} }{1-i(\omega-\omega_-)/\gamma_-} + \frac{ \e{-i\frac{4\pi}{3}} }{1-i(\omega-\omega_+)/\gamma_+} \right) \right|^2\\
T_{1\to3} &=  \left|  \frac{2}{3} \left( \frac{ \e{-i\frac{4\pi}{3}} }{1-i(\omega-\omega_-)/\gamma_-} + \frac{ \e{-i\frac{2\pi}{3}} }{1-i(\omega-\omega_+)/\gamma_+} \right) \right|^2
\end{align}
\end{subequations}
where $\omega_\pm$ and $-\gamma_\pm$ are the real and imaginary
parts of the natural mode frequencies $\omega_n$ associated with
$n=\pm 1$. Using the approximate expression \eqref{E:approx_omega_n}
for $\omega_n$, we find that for $\omega=\omega_0$ and
$\Omega=\Gamma/\sqrt{3}$ one has $T_{1\to2}=0$ and $T_{1\to3}=1$ and
thus the system behaves as an ideal circulator. Remarkably, the
energy flow in the circulator is towards the direction opposite to
that of the motion the inner cylinder.

\begin{figure}[!ht]
\centering
\includegraphics[width=.8\linewidth]{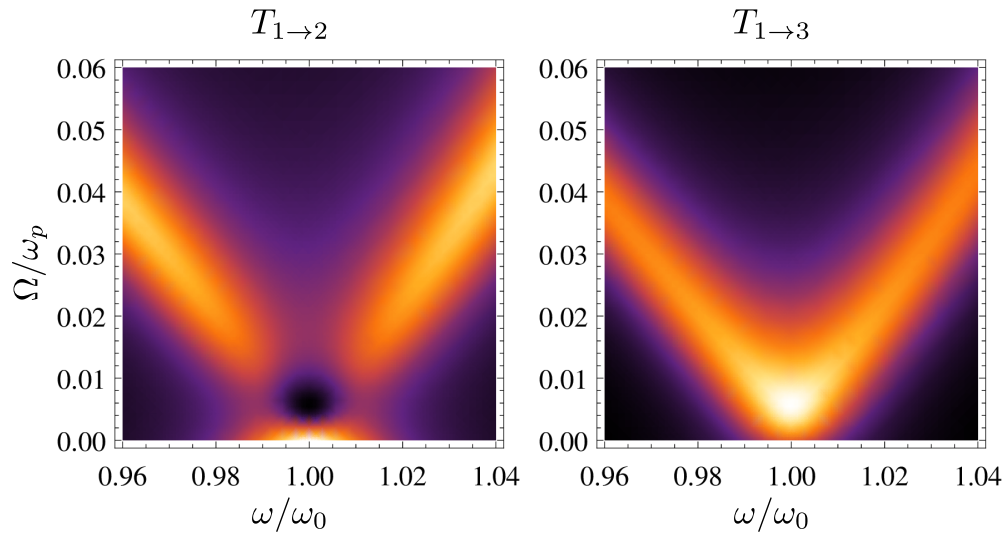}
         \caption{Density plot of the transmissivities $T_{1\to2}$ and $T_{1\to3}$
          as a function of the normalized frequency $\omega$ and of the normalized angular velocity $\Omega$.
          Here, $\omega_0$ is the resonance frequency when the walls are at rest.
          The cavity has the same structural parameters as in Fig. \ref{fig:modes_optical_circulator}.
          Brighter (darker) colours represent a stronger (weaker) transmission. }
\label{fig:transmission_density_optical_circulator}
\end{figure}

\begin{figure}[!ht]
\centering
\includegraphics[width=.6\linewidth]{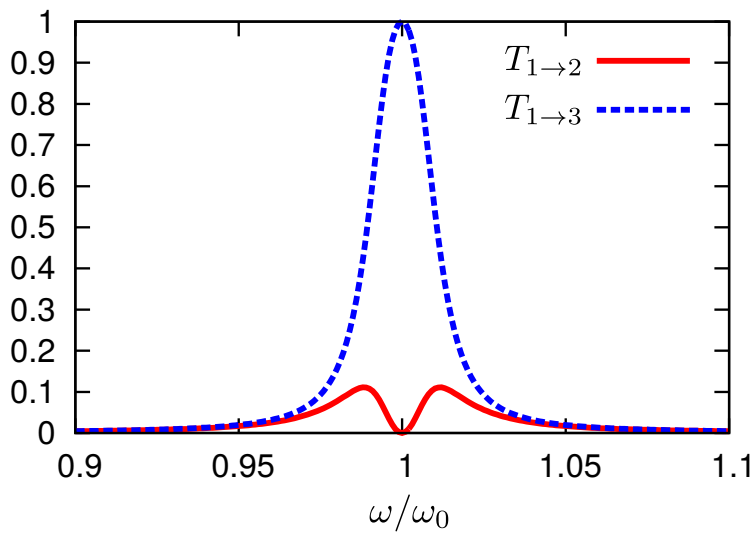}
         \caption{Transmissivities $T_{1\to2}$ and $T_{1\to3}$ as a function of the normalized
          frequency for the optimal angular velocity $\Omega = \Gamma/\sqrt{3}$ and a plasma collision frequency
           $\Gamma=0.01 \cdot \omega_p$.
           The structural parameters of the cavity are the same as in Figs. \ref{fig:modes_optical_circulator} and \ref{fig:transmission_density_optical_circulator}.}
\label{fig:transmission_optical_circulator}
\end{figure}

To confirm this result, a density plot of the transmissivities is
shown in Fig. \ref{fig:transmission_density_optical_circulator} as a
function of the normalized frequency and of the normalized angular
velocity, for the same structural parameters as in Fig.
\ref{fig:modes_optical_circulator}.
As seen in the figure, there is a region near to $\omega \approx
\omega_0$ and $\Omega \approx \Gamma/\sqrt{3} $ where simultaneously $T_{1\to2}$ goes to zero and
$T_{1\to3}$ goes to 1. In this regime, the moving cavity is strongly
non-reciprocal and behaves as an ideal circulator.
The corresponding transmission curves for an angular velocity close to the optimal angular velocity $\Omega = \Gamma/\sqrt{3} $ are represented in Fig. \ref{fig:transmission_optical_circulator} as
a function of frequency.

Interestingly, the optimal angular velocity depends only on the
level of loss in the plasmonic material, and is smaller for
resonances with high quality factor. In an optical design relying on
plasmonic materials, in the best scenario it can be about three
orders of magnitude smaller than the plasma frequency. Thus,
typically the angular velocity required for the cavity to operate as
an optical circulator lies well below the threshold associated with
electromagnetic instabilities. This puts into evidence that the
regime wherein the wave has strongly asymmetric azimuthal flows is
independent of the regime wherein the instabilities are developed.


\section{Discussion and Conclusion}

It is important to discuss the practical feasibility of the light
generator/amplifier or optical circulator studied in the previous
sections. Remarkably, the angular velocities required to obtain
optical instabilities are comparable to the metal plasma frequency,
and in the best case scenario are about three orders of magnitude
smaller for an optical circulator. To our best knowledge, the
highest angular velocities ever reached experimentally are reported
in \cite{arita_laser-induced_2013}, where a circularly polarized
laser impinging on a birefringent particle produces an angular
velocity of the order of the $\mega\hertz$. This value is far from
reaching the plasma frequencies of metals that typically are in the
UV range \cite{west_searching_2010}, or the plasma frequency of
semiconductors such as InSb that are in the \tera\hertz~ range
\cite{palik_handbook_1991, zhao_rotational_2012}. Consequently, a
direct laboratory verification of the concepts proposed in this
paper appears unfeasible with the available technology.

Yet, there may be a way to overcome these difficulties. Indeed, we
envision that the physical motion of neutral bodies can be mimicked
by an electron drift (electrons flowing on a positive ion
background) induced by a DC generator. A preliminary assessment of
this idea was reported in Ref. \cite{morgado_spontaneous_2015},
where it was demonstrated that the physics of the two systems is
rather similar in the planar case. The emergence of electromagnetic
instabilities due to an electron drift was also discussed in Refs.
\cite{riyopoulos_thz_2005,sydoruk_terahertz_2010}, where
instabilities were found at \tera\hertz~ frequencies due to the
interaction of drifting electrons with lattice waves within the same
high mobility semiconductor. The analogies between the two platforms
may offer a viable alternative to the actual motion of neutral
matter and may allow for the experimental verification of the
effects discussed in this article. These ideas will be investigated
in future work.

In summary, we studied the conditions under which a pair of
plasmonic cylinders rotating past one another develop wave
instabilities due to the hybridization of the surface plasmon
polaritons supported by the individual cylinders. The characteristic
equation for the system natural modes was found, and it was shown
that for certain cylindrical harmonics, when the angular velocity
surpasses a threshold value comparable to the plasma frequency, some
natural modes may have a positive imaginary part corresponding to a
wave amplitude growing with time. The instabilities were shown to be
robust with respect to realistic material losses. By computing the
mechanical torque acting on the cylinders, it was demonstrated that
analogous to the planar case, the wave amplification corresponds to
a conversion of kinetic energy into electromagnetic radiation and is
observed as long as the velocity of rotation is kept above the
threshold.

We also investigated the possibility of having a strongly asymmetric
light transmission relying on the rotation of the cavity walls. It
was shown that the motion of the walls induces a frequency split of
the cavity modes that results in a strong nonreciprocal behavior
that can serve to design an optical circulator. The optimal angular
frequency of rotation is determined by the quality factor of the
cavity. Cavities with larger quality factors require lower angular
velocities, and hence are the most interesting platform to
demonstrate our designs. Finally, we suggested that the behavior of
the moving walls may be mimicked by an electron drift induced by a
DC voltage generator. This concept can provide a practical roadmap
to verify and explore the proposed ideas at terahertz frequencies.

\begin{acknowledgments}
This work was partially funded by Funda\c{c}\~{a}o para Ci\^{e}ncia
e a Tecnologia under project PTDC/EEI-TEL/4543/2014.
\end{acknowledgments}


%

\end{document}